\documentclass[fleqn,usenatbib]{mnras}

\usepackage{newtxtext,newtxmath}
\usepackage[T1]{fontenc}

\DeclareRobustCommand{\VAN}[3]{#2}
\let\VANthebibliography\thebibliography
\def\thebibliography{\DeclareRobustCommand{\VAN}[3]{##3}\VANthebibliography}

\usepackage{graphicx,graphics}	% Including figure files
\usepackage{amsmath}	% Advanced maths commands
\usepackage{ulem}
\usepackage{tabularx}
\usepackage{bm,url}
\usepackage{times}
\usepackage{color}
\usepackage{float}
\usepackage{arydshln}
\usepackage{multirow}
\usepackage{xcolor}
\usepackage[switch]{lineno}
\usepackage{multicol}
\graphicspath{{./fig/}{./png/}}

\def\Rs{R_{s}}
\newcommand{\bl}{Babcock--Leighton}
\newcommand{\dr}{differential rotation}
\newcommand{\Fig}[1]{Fig.~\ref{#1}}
\newcommand{\Figs}[2]{Figs.~\ref{#1} and \ref{#2}}
\newcommand{\Eq}[1]{Eq.~(\ref{#1})}

\def\Rs{R_{s}}
\newcommand{\etasurf}{\eta_{\mathrm{surf}}}
\newcommand{\etaSCZ}{\eta_{\mathrm{SCZ}}}
\newcommand{\etaRZ}{\eta_{\mathrm{RZ}}}
\newcommand{\rBCZ}{r_{\mathrm{BCZ}}}
\newcommand{\rsurf}{r_{\mathrm{surf}}}

\def\mc{meridional circulation}
\title[Cycle variability and grand minima in Sun-like stars]{
Dynamo modelling for cycle variability and occurrence of grand minima in Sun-like stars: Rotation rate dependence
}

\author[V. Vashishth]{
Vindya Vashishth$^{1}$\thanks{E-mail: vindyavashishth.rs.phy19@iitbhu.ac.in}
 and Bidya Binay Karak$^{1}$\thanks{E-mail: karak.phy@iitbhu.ac.in}
 and Leonid Kitchatinov$^{2}$
\\
$^{1}$Department of Physics, Indian Institute of Technology (BHU), Varanasi 221005, India\\
$^{2}$Institute of Solar-Terrestrial Physics SB RAS, Lermontov Str. 126A, Irkutsk 664033, Russia
}

\date{Accepted XXX. Received YYY; in original form ZZZ}

\pubyear{2023}

\begin{document}
\label{firstpage}
\pagerange{\pageref{firstpage}--\pageref{lastpage}}
\maketitle

\begin{abstract}
Like the solar cycle, stellar activity cycles are also irregular. Observations reveal that rapidly rotating (young) Sun-like stars exhibit a high level of activity with no Maunder-like grand minima and rarely display smooth regular activity cycles. On the other hand, slowly rotating old stars like the Sun  have low activity levels and smooth cycles with occasional grand minima. We, for the first time, try to model these observational trends using flux transport dynamo models. Following previous works, we build kinematic dynamo models of one solar mass star with different rotation rates. Differential rotation and meridional circulation are specified with a mean-field hydrodynamic model. We include stochastic fluctuations in the Babcock--Leighton source of the poloidal field to capture the inherent fluctuations in the stellar convection. Based on extensive simulations, we find that rapidly rotating stars produce highly irregular cycles with strong magnetic fields and rarely produce Maunder-like grand minima, whereas the slowly-rotating stars (with a rotation period of 10 days and longer) produce smooth cycles of weaker strength, long-term modulation in the amplitude, and occasional extended grand minima. The average duration and the frequency of grand minima increase with decreasing rotation rate. These results can be understood as the tendency of less supercritical dynamo in slower rotating stars to be more prone to produce extended grand minima. 
\end{abstract}


%%%%%%%%%%%%%%%%%%%%%%%%%%%%%%%%%%%%%%%%%%%%%%%%%%

%%%%%%%%%%%%%%%%% BODY OF PAPER %%%%%%%%%%%%%%%%%%

\section{Introduction}

Sun is not the only star that has a complex and variable
magnetic field.
Chromospheric emission of Ca II H \& K of many stars with spectral types from early F to M observed since 1966 through the HK Project of Mount Wilson Observatory (MWO) revealed the magnetic cycles \citep{Baliu95}.
%and also confirm the strong activity cycle period relation with the rotation rate of the star.
Other recent observations from the coronal X-ray emission \citep{wright11, WD16} and the magnetic field through the Zeeman-Doppler Imaging (ZDI) \citep{donati92, Vidotto_EA_14} provide further evidence for magnetic activity in many stars studied earlier using the Ca~II H \& K emission for the longer duration.

Stellar magnetic activity is observed to be largely controlled by rotation. 
The more rapidly a star rotates, the more active it is \citep{Skumanich72, R84}. \citet{Noyes84a} and \citet{WD16} gave the activity-rotation relation using Ca II H \& K and X-ray emissions, respectively. They showed that the activity increases with the  increase in rotation rate (or decrease in rotation period) for slow and moderate rotators, and then the activity tends to saturate for the fast-rotating stars. 

While most of the earlier observations focused on the activity-rotation relation, some observations find a trend between the cycle duration and the rotation period of different sun-like stars \citep{Noyes84b, SoonBliunas1994}. 
%\citep{Noyes84b,Brown08,SoonBliunas1994}. 
Using the chromospheric activity from HARPS (High Accuracy Radial velocity Planet Searcher) and MWO data of 4454
cool stars, \citet{BoroSaikia18} showed that the trend of the cycle period with the rotation period for the fast rotators is different from the slowly-rotating stars. As the rotation period increases, the cycle period somewhat decreases for the rapidly rotating stars and increases for the slowly-rotating stars. The trend is, however, quite complicated for fast rotators. 

Our Sun shows the magnetic cycle of 22 years period (11 years in strength) with amplitude varying somewhat smoothly in the long-term (beyond the 11-year period)
and show occasional extended periods of weaker activity, the grand minima, e.g., Maunder minimum \citep{Uso17}. 
Different stars, in contrast, show a wide range of variability in the magnetic cycles.
\citet{Baliu95} observed a smoother variability and occasional grand minima in the magnetic cycles of the slowly rotating (old) stars. 
On the other hand, they observed much irregular activity and no grand minima in the rapidly rotating young stars. 
\citet{Olah16, BoroSaikia18, garg19} also produced similar evidences using additional data.
Recently, \citet{anna2022} claimed that a K2V star, HD 166620, has entered  into a grand minimum phase, and interestingly, it is a slowly rotating star. Also, \citet{Shah18} suggests that HD 4915 is a possible Maunder minimum candidate, although its rotation period still needs to be confirmed.

A large-scale dynamo, powered by the helical convection and \dr, is responsible for the generation of the magnetic cycle in the Sun
\citep{Pa55}. 
As the other sun-like stars have convection zones (CZs) in their outer layers,
it is natural to expect that these stars also support dynamo action through which the stellar magnetic cycles are maintained. 
Some of the stellar cycles (e.g., HD 10476, HD 16160, etc.) are so similar to the solar cycle \citep[in terms of regular cyclic variation and obeying the Waldmeier effect, which says that strong cycles rise faster than the weaker ones;][]{garg19} that it suggests a similar dynamo operating in sun and other sun-like stars \citep[also see][for a theoretical argument behind this expectation]{jeffers22}.
The motivation of our work is to extract the dependency on the rotation rate of the sun-like stars for their cycle variability and the occurrence of the grand minima using dynamo models.

Recently, the \bl\ mechanism \citep{Ba61, Leighton69}, in which the tilted bipolar magnetic regions produce poloidal field in the sun, 
has received strong observational supports 
\citep{Das10, KO11b,Priy14,CS15}.
Including this process for the generation of the poloidal field,
the \bl\ type dynamo models have produced great successes in providing many observational features of the solar magnetic cycle, including the grand minima \citep[e.g.,][]{CK12, OK13, Pas14, KM17, CS17, LC17, Fadil17, BKC22}.

In the past, 
\bl\ dynamo model has also been used to study the stellar magnetic cycles. For 
instance, \citet{NM07} employed a time-delay dynamo model to investigate the relationship between the magnetic field and cycle period with the dynamo number.
\citet{JBB10} utilized a kinematic Babcock-Leighton dynamo model to observe that the cycle period increases as the rotation rate increases, unless the meridional flow speed is assumed to increase with the rotation rate, which contradicts theoretical results \citep{Miesch05, Brown08}. Later, by employing the non-local and distributed $\alpha$~effects in non-linear $\alpha^2$ dynamo models for moderate to slowly rotating stars,  \citet{Pipin15} find some agreement of the cycle period vs rotation period with observation.
\cite{KKC14} constructed the \bl\ type flux transport dynamo model for sun-like stars with different rotation periods by including differential rotation and \mc\ from corresponding hydrodynamical models of \citet{KO11b}. 
They managed to reproduce the activity-rotation relation correctly but again failed to reproduce the cycle period vs rotation rate relation. Recently, \citet{Hazra19} performed simulations using the same model but by including a radial  pumping near the surface of the stars. 
They found an increasing trend of the cycle period with an increase in 
the rotation period for the slowly-rotating stars, and a decreasing trend in 
the cycle period for the rapidly rotating stars; also see \citet{Docao11} who included latitudinal pumping and found some agreement with observations. 
\citet{K22} studied the stellar activity cycles using a \bl\ type dynamo model and showed a 
strong temperature dependence on the cycle period.
\citet{KTV20} and \citet{NB22} applied mean-field models in different stars and addressed the effect of anti-solar differential rotation \citep[which naturally arises in the high-Rossby number convection, e.g.,][]{gastin, Brun17, KMB18b}  on the polarity reversal and the strength of magnetic field in slowly-rotating stars. 
However, to the best of our knowledge, no previous study was performed to explore the cycle variability and grand minima in stars. 

In recent years, global MHD convection simulations have produced some 
exciting results of the stellar magnetic cycles \citep{Kar15, ABMT15, Kap16, Strugarek18, viviani19, Brun22}.  \citet{War18} analyzed how the magnetic cycle period changes as a function of the Rossby number. \cite{viv18} studied the simulations of different stars and showed the transition of the magnetic field from axi- to non-axisymmetric field configuration at around 1.8 times the solar rotation rate, where the differential rotation changes from solar to anti-solar. 
However, these MHD simulations are not capable of reproducing some basic features of the solar cycle (e.g., the 11-year periodicity with regular reversal, equatorward migration of toroidal field at low latitude, poleward migration of surface radial field, largely dipolar field) and the correct flow (particularly the observed amplitude of the convective flow)  robustly, 
there lies the uncertainty of whether these results hold in stars. Being computationally expensive, these simulations were not applied to study the long-term variability of stellar cycles; however, see \citet{PC14, ABMT15, Kap16}, who have performed simulations for several cycles which may be used for long-term studies.

In this paper, we apply the models of \cite{KKC14} and \citet{Hazra19} to study the irregularities of the stellar cycle, 
in particular, how the variability and the frequency of grand minima change with the stellar rotation. 
For this, we shall include the stochastic fluctuations to capture the inherent randomness  in the stellar convection \citep{C92} as seen in the form of noise in the flux emergence and the tilts of BMRs around Joy’s law \citep{Das10, SK12, MNL14, Wang15, Arlt16, Jha20} (Section~\ref{sec:model}).
We shall see that our  models
produce a strong magnetic activity and highly irregular cycles in rapidly rotating stars 
and, on the contrary, a weak magnetic activity and more regular cycles in slowly-rotating stars (Section~\ref{sec:results-1}). 
Maunder-like extended grand minima are only produced in slowly-rotating stars (with rotation period of 10 days and longer), and the frequency of occurrence of these events increase with the increase in the rotation period of the star (Section~\ref{sec:variability}).\\

\section{Model}
\label{sec:model}
We build our model based on \citet{KKC14} where the following equations for the axisymmetric magnetic field are evolved, 

\begin{equation}
~~~~~~~~~\frac{\partial A}{\partial t} + \frac{1}{s}({\bf v_p}.\nabla)(s A)
= \eta \left( \nabla^2 - \frac{1}{s^2} \right) A + S(r, \theta; B),
\label{eq:pol}
\end{equation}
\begin{eqnarray}
\frac{\partial B}{\partial t}
+ \frac{1}{r} \left[ \frac{\partial}{\partial r}
(r v_r B) + \frac{\partial}{\partial \theta}(v_{\theta} B) \right]
= \eta \left( \nabla^2 - \frac{1}{s^2} \right) B \nonumber \\
+\ s({\bf B_p} \cdot {\bf \nabla})\Omega + \frac{1}{r} \frac{d\eta}{dr} \frac{\partial} {\partial r} (rB),~~~~~~~~~~~~~~~~~~~~~~~~~~~~~
\label{eq:tor}
\end{eqnarray}\\
where 
${\bf B_p} = \nabla \times [ A(r, \theta) {\bf e}_{\phi}]$ is the poloidal component of the magnetic field and
$B(r, \theta)$ is the toroidal component,
$s = r \sin \theta$, ${\bf v_p} = v_r { \hat r} + v_{\theta} {\bf \hat \theta}$ is the meridional circulation, $\Omega$ is the angular velocity. 
While $\Omega$ is well-measured in the whole CZ of the sun, the meridional flow is only constrained in the near-surface layer of the Sun. Recent helioseismic studies for the deep meridional circulation indicate a single-cell flow in the solar convection zone 
\citep{RA15, gizon20}. Observations for other stars, on the other hand, are limited to the surface differential rotation only. Global MHD simulations for the sun-like stars provide differential rotation, which often shows transition from solar to anti-solar profile with increasing Rossby number near the solar value, and the meridional flow, which is multi-cellular and time varying \citep[e.g.,][]{FM15, Kar15, viviani19}.
Therefore, for ${\bf v_p}$ and $\Omega$, we use the data from a mean-field hydrodynamic model of \citet{KO11b}.
This numerical model 
jointly solves the mean-field equations for the angular velocity, meridional flow, and heat transport in a spherical layer of a  stellar convection zone. The model produces the solar-type \dr\ as a consequence of angular momentum fluxes. The one-cell per hemisphere meridional flow predicted by the model for the sun agrees with the recent seismological detection \citep{RA15, gizon20}. Also, the computed dependence of \dr\ on stellar rotation rate and spectral type \citep{Kit_Ole_12DR} is in at least qualitative agreement with observations by \citet{Barnes_EA_05} and \citet{Bal_Abedi_16}.
The model, however, does not produce the tachocline 
self-consistently; rather, it has the lower boundary at $0.72\Rs$ ($\Rs$ being the stellar radius).  Therefore, the tachocline in our dynamo model is formed by smoothly varying the $\Omega_{\rm model}(r,
\theta)$ from the differential rotation model at $r = 0.72\Rs$ to the value of the rotation rate at the core $\Omega_{\rm core}$ in the following way:
\begin{eqnarray}
    \Omega (r,\theta) = \Omega_{\rm model}(r,\theta) + \frac{1}{2} \left[\Omega_{\rm core} - \Omega_{\rm model}(0.72\Rs, \theta)\right] \\ \nonumber
     \times \left[1-\mathrm{erf}\left(\frac{r-0.7\Rs}{0.02\Rs}\right)\right].
\end{eqnarray}
The $\eta$ is the turbulent magnetic diffusivity and is taken as a function of $r$ alone, having the following form:
\begin{eqnarray}
\eta(r) = \etaRZ + \frac{\etaSCZ}{2}\left[1 + \mathrm{erf} \left(\frac{r - \rBCZ}
{d_t}\right) \right]\nonumber \\
+\frac{\etasurf}{2}\left[1 + \mathrm{erf} \left(\frac{r - \rsurf}
{d_2}\right) \right]
\label{eq:eta}
\end{eqnarray}\\
with $\rBCZ=0.7 \Rs$, $d_t=0.015 \Rs$, $d_2=0.05 \Rs$, $\rsurf = 0.95 \Rs$,
$\etaRZ$, $\etaSCZ$, and $\etasurf$ represent the diffusivities, at the inner boundary, within CZ, and at the surface respectively, having the values as $\etaRZ = 5 \times 10^8$ cm$^2$~s$^{-1}$, $\etaSCZ = 5 \times 10^{10}$ cm$^2$~s$^{-1}$, and
$\etasurf = 2\times10^{12}$ cm$^2$ s$^{-1}$.
The diffusivity profile of \Eq{eq:eta} approaches $\etaRZ$ at the inner boundary of $0.6 \Rs$, remains at  
$\etaSCZ$ in the bulk of CZ
and increases to $\etasurf$ at the surface \citep[see figure 5 in][]{KKC14}. 
We have ignored the change of diffusivity with the rotation rate and the magnetic field in our study due to its limited knowledge in estimating its value in different stars.

The term $S(r, \theta; B)$ is the source for the generation of poloidal field which captures the Babcock--Leighton mechanism in our axisymmetric model, and it is given by,
\begin{equation}
~~~~~~~~~~~~~~~~~ S(r, \theta; B) = \frac{ \alpha_0 ~ \alpha(r,\theta)}{1 + \left( \overline{B} (r_t,\theta)/B_0 \right)^2} \overline{B}(r_t,\theta),
\label{source}
\end{equation}
where $\alpha_0$ is 
the strength of \bl\ process, 
$\overline{B}(r_t,\theta)$ is the toroidal field at latitude $\theta$
averaged over the whole tachocline $r = 0.685 \Rs$ to $r=0.715 \Rs$. 
From \Eq{source} we observe that when the magnetic field becomes comparable to $B_0$ (the saturation field strength), the nonlinearity becomes important and the field eventually tends to hover around $B_0$. Therefore, everywhere in our study, we measure the magnetic field in the unit of $B_0$.
While in the traditional $\alpha$ effect based on the helical convection, above $\alpha$ quenching is obvious, the \bl\ $\alpha$ also experiences a
quenching due to the fact that BMRs with strong fields rise quickly and Coriolis force gets less time to induce tilt and the strong magnetic field also gives more tension which causes less tilt \citep{DC93, FFM94}. 
Observations indeed find some evidence of tilt quenching \citep{Jha20}. Furthermore, the latitudinal variation of BMRs (stronger cycles produce BMRs at higher latitudes) gives rise to quenching in the \bl\ process \citep{J20, Kar20}.

We have limited knowledge about the \bl\ process in other stars and thus we are not sure of how the strength of this process changes with the rotation of the star. On theoretical grounds, we expect the tilt of bipolar magnetic regions to increase with the increase of the rotation rate of stars \citep{DC93, KO15}. However, there is an opposing effect that arises due to the fact that with the increase of rotation rate, the latitudes of BMR emergences are expected to shift to higher latitudes \citep{SS92} and higher latitudes BMRs are less efficient in producing polar field \citep{JCS14, Kar20}. 
In our study, the strength of \bl\ process $\alpha_0$, is chosen to depend on the rotation in the following way,
\begin{eqnarray}
 \alpha_0 = \alpha_{0,s} \frac{P_\odot}{P_{\rm rot}},
 \label{eq:alpha0I}
\end{eqnarray}
where $\alpha_{0,s}$ is the value of $\alpha_0$
for the solar case and $P_\odot$ and $P_{\rm rot}$ are the rotation period of Sun and the star, respectively.

The \bl\ process includes considerable randomness, primarily due to irregular variations in the tilts and emergence rates of the bipolar active regions \citep{JCS14}. 
Therefore, we include fluctuations in the $\alpha$ appearing in \Eq{eq:alpha0I} in the following way. 
%$$\alpha_{0,s} = \alpha_{0,s} X_n, $$  
\begin{eqnarray}
\alpha_{0,s} = \alpha_{0,s} X_n,
\end{eqnarray}
where $X_n$ is the Gaussian random number with a mean of 
unity and standard deviation ($\sigma$) of 2.67. This value of $\sigma$ is inspired by the study of \citet{OKC13}, who computed the fluctuations in the 
Babcock--Leighton process by estimating the contribution of the sunspot group to the polar field using the data of Royal Greenwich, Kodaikanal and
Mount Wilson Observatories.
We keep the value of $\sigma$ same throughout all the simulations presented in this paper.
In our models, the value of $\alpha_0$ is updated randomly after a certain time step which we take to be one month.

It may be noted that other model parameters like eddy diffusivity or turbulent pumping can also include fluctuations due to randomness inherent to turbulent stellar convection. However, the fluctuations are relatively small and less consequential compared to the fluctuations in the \bl\ mechanism. In particular, fluctuations in the angular momentum fluxes produce only small variations in the differential rotation and moderate fluctuations in the meridional flow \citep{Rempel_05_Random_Lambda}. Thus, these fluctuations are neglected. The differential rotation and meridional flow are steady and equator-symmetric in our model.

\subsection{Model I}
In this case, we use the same model as given in \citet{KKC14} except 
the value of the strength of the \bl\ $\alpha$, and we add fluctuations in the \bl\ process. 
In this model, the value of $\alpha_0$ is taken as $0.9$ cm~s$^{-1}$ (instead of $1.6$ cm~s$^{-1}$ which was used in \citet{KKC14}).

As in \bl\ models, $\alpha$ captures the average effect of the decay of tilted BMRs, it must be non-zero only near the surface, and it must have $\cos\theta$ dependence due to the angular dependence of the Coriolis force which is the possible cause of BMR tilt. However, to suppress the poloidal field generation in high latitudes (as BMRs do not appear in high latitudes), a $\sin\theta$ factor is also introduced \citep[see, e.g.,][] {DC99}. Therefore in this model, the
profile of $\alpha$ is given by,
\begin{equation}
\alpha(r,\theta)=\frac{1}{4}\left[1+\mathrm{erf}\left(\frac{r-r_4}{d_4}\right)\right]\left[1-\mathrm{erf}\left(\frac{r-r_5}{d_5}\right)\right] \sin\theta\cos\theta
\label{alpha}
\end{equation}
with $r_4=0.95 \Rs$, $r_5= \Rs$, $d_4=0.05 \Rs$, $d_5=0.01 \Rs$.

\subsection{Model II}
In the previous model (Model~I), one hemisphere of the sun was
used to study the dynamo, for which 
a dipolar boundary condition was imposed at the equator. 
This did not allow us to observe the magnetic field configuration across the equator and the hemispheric asymmetry of the magnetic  field, which are very relevant for the solar and stellar observations \citep{DBH12}. Therefore, in Model~II, we extend the same Model~I to the full sphere of the sun, and we include the fluctuations separately in the two hemispheres.
Other than extending Model~I to the full sphere and thus eliminating the equatorial boundary condition, no other changes are made.% in Model~II. 

\subsection{Model III}

Finally, we take Model~III which is same as Model~II but at increased diffusivity and added radial magnetic pumping. 
This inclusion of pumping is inspired by \citet{Hazra19}, who found some agreement of the cycle period vs rotation trend with observations. 
It was realized that a downward magnetic pumping helps to make the magnetic field radial near the surface and reduce the toroidal flux loss through the surface, making the dynamo model in accordance with the surface flux transport models and observations \citep{Ca12}. The near-surface pumping also helps the dynamo to operate at a high diffusivity range consistent with the mixing-length theory \citep{KO12, KC16, KM17}, and facilitates the model to recover from the Maunder-like extended grand minima \citep{KM18}. 
The pumping has the following form:
\begin{equation}
~~~~~~~~~~~~~~~~~~~~~~~~~~~ \gamma = - \gamma_0 \left[1 + \mathrm{erf} \left(\frac{r - 0.9\Rs}
{0.02\Rs}\right) \right],
\label{pumping}
\end{equation}
where the amplitude of the radial magnetic pumping is given by $\gamma_0$ which is 24 m\,s$^{-1}$ in all the stars.

We note that we do not use the 
exact same model of \citet{Hazra19} because, in that model, when we include fluctuations, even the solar case does not produce the dipolar field as seen in the observations.
Therefore, to obtain the dipolar field, we reduce the diffusivity for the bulk of the CZ by taking  the following parameters (\Eq{eq:eta}):
$\etaSCZ = 3 \times 10^{11}$ cm$^2$~s$^{-1}$, and
$\etasurf = 3\times10^{12}$ cm$^2$ s$^{-1}$. 
We note that with these parameters, the diffusivity in the whole CZ is about six times stronger than that used in Models I and II. 
%-----------------------------------------------

The $\alpha$ profile used in this model is given by,
\begin{eqnarray}
\alpha(r,\theta)=\frac{1}{2}\left[1+\mathrm{erf}\left(\frac{r-\rsurf}{d}\right)\right] \sin ^{2}\theta\cos\theta,
\label{alpha_model3}
\end{eqnarray}
where $d= 0.01\Rs$. 
The $\alpha_0$ has the same form  (\Eq{eq:alpha0I}) as in Model~I, except in this case, $\alpha_{0,s} = 4$ cm~s$^{-1}$ and fluctuations in this model are included separately in the two hemispheres. We note that above $\alpha$ in \Eq{alpha_model3} has a $\sin ^{2}\theta\cos\theta$ dependence instead of $ \sin\theta\cos\theta$ as used in Models I-II to make the $\alpha$ effect strong (weaker) in low (high) latitudes. Also, the radial 
extent of this $\alpha$ is a bit wider than that used in Models I-II.  

After specifying all the parameters, we solve the above equations \ref{eq:pol} and 
\ref{eq:tor} numerically in the SCZ with the radial extent of $0.55 \Rs$ to $\Rs$ and the following boundary conditions. We take, at the lower boundary: $A = B = 0$, at the top (surface) layer: $B = B_\theta =  0 $ (i.e., radial 
boundary condition),  at poles: $B = A = 0$ (i.e., no singularity) and at the equator for Model~I: $B=0 = \partial A / \partial \theta = 0$ (dipolar condition).
Simulations are performed in $129\times129$ grid points in radial and latitudinal directions.

\section{Results and Discussions}
\label{sec:results}
We initialized our dynamo simulations for stars of solar mass with rotation periods of 1, 3, 7, 10, 15, 20, 25.38 (Sun), and 30 days, 
respectively, by computing differential rotation and meridional circulation from the mean-field hydrodynamic model of \citet{KO11b}. 
The following sections discuss the various aspects of magnetic cycles obtained from all three models.

\subsection{Magnetic field morphology}
\label{sec:results-1}
\begin{figure}
\centering
\includegraphics[width=1.1\columnwidth]{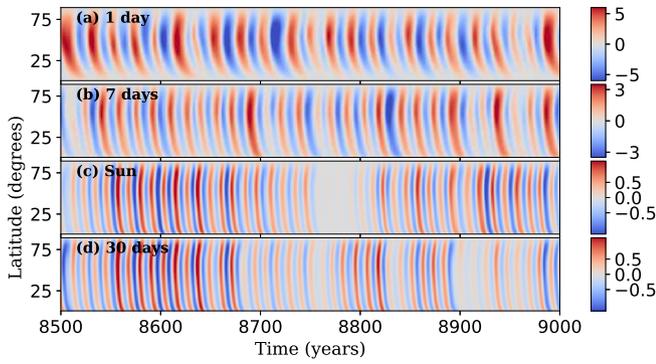}
\vspace{-0.2in}
\caption{Time–latitude plots of toroidal field at $r = 0.71 R_s$ (in the unit of $B_0$)
  for different stars with rotation period of (a) 1 day, (b) 7 days, (c) the solar value ( i.e., 25.38 days), and (d) 30 days for Model I.}
  \label{fig:quarter_bfly}
\end{figure}

\begin{figure}
\centering
 \includegraphics[width=1.08\linewidth]{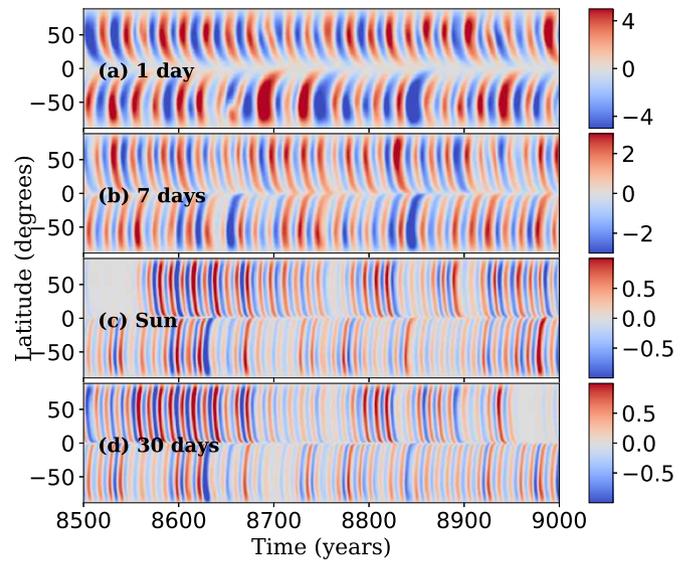}  
\vspace{-0.2in}
\caption{Same as \Fig{fig:quarter_bfly}, but for  Model II}
\label{fig:fullhem_bfly}
\end{figure}

\begin{figure}
\centering
\includegraphics[width=1.05\linewidth]{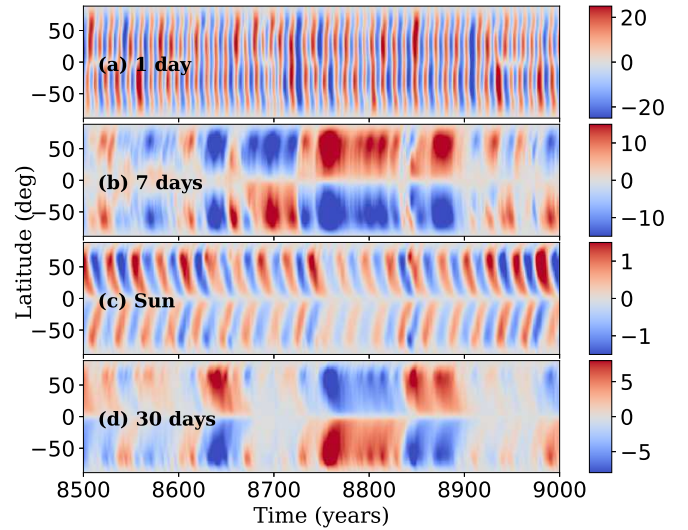}
\vspace{-0.2in}
\caption{Same as \Fig{fig:quarter_bfly}, but from Model III.}
\label{fig:hazra_bfly}
\end{figure}

%polarity reversal
\Fig{fig:quarter_bfly} depicts the butterfly diagrams of the toroidal field at the base of CZ for the rotation periods 1, 7, 25.38, and 30 days from Model~I.
After analyzing these panels,
we find the regular polarity reversal. However,  in Model~II 
(see \Fig{fig:fullhem_bfly}) the magnetic cycles are 
a bit irregular. 
We observe a strong hemispheric asymmetry in the magnetic field. 
Sometimes the magnetic field in 
one hemisphere is largely suppressed or enhanced. 
Hence, in Model~II, the polarity reversal is not regular. Finally, for Model~III (\Fig{fig:hazra_bfly}), we  observe that for the rapidly rotating case and the Sun, regular reversal is seen, but for 
intermediate and slow rotators, the polarity reversal is not regular.
The magnetic field distribution for the rotation period of 7 and 30 days show extended cycles. However, for the 7 days case, the magnetic field is largely quadrupolar, while for the 30 days case, it is largely bipolar. We observe that the magnetic field distribution at higher radial layers is largely different.
We also note that for the slowly-rotating stars with a rotation period of $\geq$ 10 days, the magnetic field is largely dipolar.
In contrast, for the rotation period of 7 days and less, the parity is changed to largely quadrupolar, although there are sometimes when the parity remains dipolar.

In all the models for the slowly-rotating stars, 
we observe an equatorward migration of the toroidal field at the low latitudes as a consequence of the 
transport by the equatorward meridional circulation 
\citep[Figure 3 of][]{KKC14}.
However, for the rapidly rotating stars, we find a weak poleward migration of the field in high latitudes, particularly see \Fig{fig:hazra_bfly}(a).
This poleward migration is due to the diffusion of the field from the mid-latitude where the toroidal field generation is strongest \citep[also see Fig.\ 6 \& 10 in][]{Hazra19}.
Moreover, the meridional flow is much weaker in rapidly rotating stars.

One obvious feature in these simulations is that the magnetic field becomes strong in fast-rotating stars. This happens because the strength of $\alpha$ increases with the rotation rate of the star. 
This increase in the magnetic field is congruous with the observations \citep{Noyes84a, wright11}.
While in observations, the magnetic field is 
saturated in rapidly rotating stars, our model
always produces an increasing trend with the rotation rate. This is because, in our model, the latitude of  operation of the  \bl\ process (the band of BMR emergences) is fixed in all stars, while in observations, it increases. Thus, the generation of poloidal field is less efficient in rapidly rotating stars \citep{KO15}.  

These time-latitude plots also give a hint about the variability observed in different stars. 
Slowly-rotating stars seem to produce more long-term modulation in their cycles, including extended episodes of weaker magnetic field. 
In contrast, fast rotators generate less modulation.
This result is in agreement with observations.
The root cause for such behavior is that the slowly-rotating stars have a small dynamo number. Due to this, if the magnetic cycle gets weaker sometimes, then it would take a long time to grow the field. Therefore, we see a long-term modulation in the slowly-rotating stars. But for the fast rotators, the cycle recovers its strength quickly after getting into the weak phase. 
In fast rotators, the dynamo number is high so the growth rate is very high. This trend is also explained in \citet{KKV21} and \citet{Vindya21}. The results are in accordance with the observations as well \citep{Baliu95}. A detailed discussion of the cycle variability is made in Section~\ref{sec:variability}.

\subsection{Cycle duration vs rotation period }
\label{sec:results-2}
\begin{figure}
\centering
\includegraphics[scale=0.35]{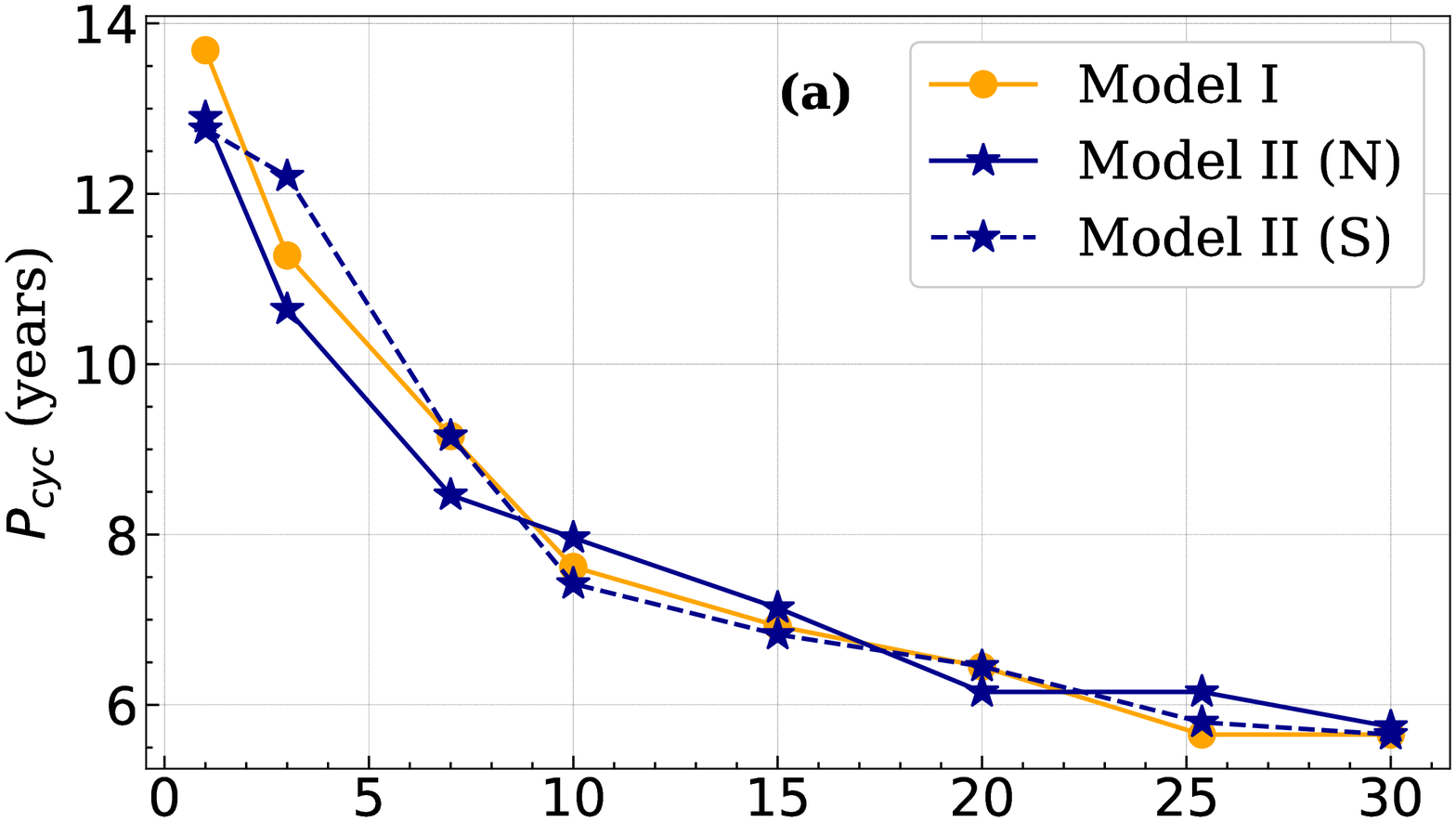}
\includegraphics[scale=0.35]{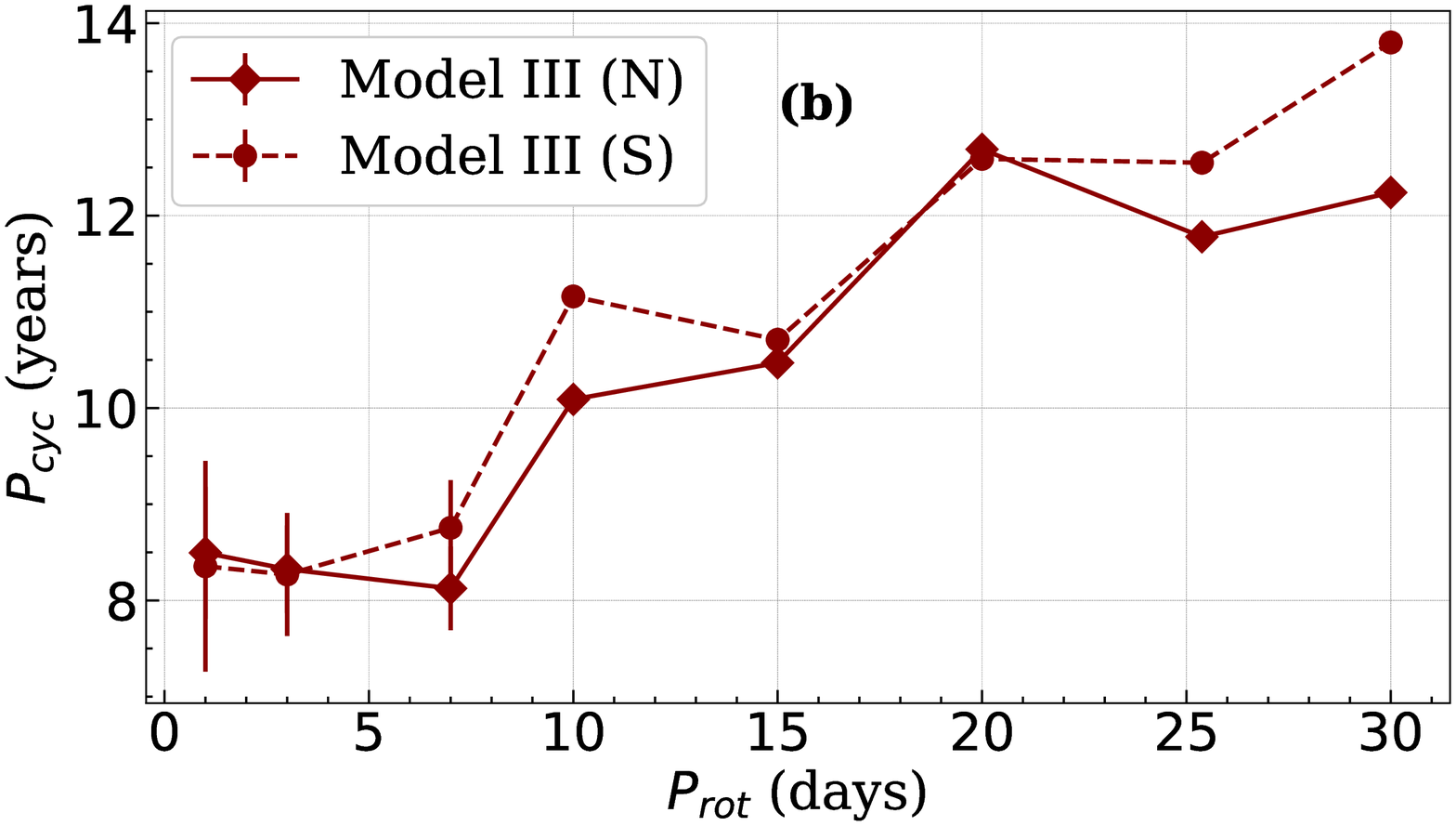}
\vspace{-0.3in}
\caption{Variations of the activity cycle period ($P_{\rm cyc}$ in years) with rotation period ($P_{\rm rot}$ in days) for (a) Models I (filled circles) and II (asterisks) and (b) Model III (solid and dashed lines are for northern and southern hemispheres, respectively).}
\label{fig:Pcy_Prot}
\end{figure}

We now compute the cycle periods for all three models. This is done by determining the peak of the Fourier power spectrum of the time series of the toroidal field over the tachocline for both the northern and southern hemispheres separately. 
However, due to the irregular nature of the cycles in Model~III, we fail to identify a prominent peak in the power spectrum and hence cannot identify the dominant cycle period for the stars having a rotation period of less than 10 days. Fortunately, we are able to find a range in which the cycle periods of these stars could lie.
The computed cycle periods for all the models are listed in Table \ref{tab:table1}, and the variations with the rotation rate are shown in \Fig{fig:Pcy_Prot}.
This figure infers an increasing trend of the cycle period with the stellar rotation rate for both Models I and II.
This increasing trend is quicker in the fast-rotating stars and milder in the slowly-rotating stars.
This happens because, as the rotation period decreases (or rotation rate increases), the meridional flow becomes weaker (although the flow speed increases in the thin layers near the top and bottom boundaries). 

\begin{table*}
\centering
\caption{Summary of simulations. Here, $P_{\rm rot}$ is the rotation period of the star in days, and $P_{\rm cyc}$ is the mean magnetic cycle period in each hemisphere. For each model, the number of grand minima and parity are computed from 
the surface radial and the toroidal fields at the base of CZ,
which are separated by a comma.}

\begin{tabular}{|c|ccc|ccc|cc|}
\hline
$P_{\rm rot}$ (d)  &  \multicolumn{3}{c |}{$P_{\rm cyc}$ (yr)} & \multicolumn{3}{c |}{No. of grand minima} & \multicolumn{2}{c |}{Parity} \\

\hline
&  Model I  &  Model II (N, S) & Model III (N, S) & Model I & Model II & Model III & Model  II & Model III \\ [1ex]
\hline
1 & 13.68 & 12.89, 12.76 & 7.81--9.18, 7.26--9.45 & 0, 0 & 0, 0 & 0, 0  &  0.013, $-0.053$ & 0.434, 0.410\\
3  & 11.27 & 10.64, 12.25 & 7.87--8.78, 7.63--8.91 & 0, 0 & 0, 0 & 0, 0 &  0.003, $-0.043$ & $-0.726$, $-0.742$\\
7 &  9.15 & 8.46, 9.15 & 7.69--8.56, 8.26--9.25 & 0, 0 & 0, 0 & 0, 0 & $-0.015$, $-0.079$ & $-0.685$, $-0.676$\\
10 & 7.62 & 7.96, 7.42 & 10.09, 11.16 & 2, 4 & 0, 0 & 2, 2 &  0.009, $-0.032$ & $-0.786$, $-0.788$\\
15 &  6.92 & 7.13, 6.82 & 10.47, 10.71 &  6, 15 & 2, 4 & 3, 4 &  0.001, $-0.038$ & $-0.844$, $-0.846$\\
20 & 6.44 & 6.15, 6.45 & 12.69, 12.59 &  17, 22 & 6, 12 & 6, 8 &  0.009, $-0.011$ & $-0.831$, $-0.830$\\
25.38 (Sun) & 5.65 & 6.15, 5.79 & 11.78, 12.55 & 32, 40 & 16,  21 & 8, 12 & $-0.117$, $-0.194$ & $-0.699$, $-0.830$\\  
30 & 5.65 & 5.74, 5.65 & 12.24, 13.80 & 36, 41 & 19, 24 & 10,16 & $-0.138$, $-0.201$ & $-0.889$, $-0.888$\\
\hline
\end{tabular}
\label{tab:table1}
\end{table*}

Although these two models reproduce various stellar observations, they fail to reproduce the magnetic cycle period vs. rotation trend correctly for the slowly-rotating stars. 
Limited observations  \citep{BoroSaikia18} seem to show a rapid increase in the cycle period with the increase of the rotation rate for fast-rotating stars. This is consistent with the trend found in our Models I-II. 
However, the observed data for slow rotators show an increasing trend of the activity cycle period with the increase in rotation period, 
which is opposite to the findings 
in Models I-II.
One way to resolve this discrepancy is to include 
radial magnetic pumping in the stellar CZs.
\citet{Hazra19}, after including the 
pumping, got a trend somewhat similar to the observations.
In our Model III, after including radial pumping, we also got the cycle-rotation period trend closer to the observations. 
It is also possible that the observed trend is a consequence of strong decrease of the cycle period with stellar effective temperature and on average faster rotation of hotter stars \citep{K22}.
A different trend of the cycle period
with the rotation period found in Model III is due to the operation of the dynamo in a pumping-dominated regime. When strong downward magnetic pumping is included in this model, the diffusion of the magnetic field across the surface becomes negligible and then the dynamo allows it to operate at a low $\alpha$ \citep{KC16}. Lower the $\alpha$ longer is the cycle period. We can see from \Fig{fig:Pcy_Prot} that at 30 days rotation period, while Models I-II were producing a cycle period of 6 years, Model III produced a much longer period of 13 years. Then with the decrease of the rotation period, the $\alpha$ becomes stronger and thus the poloidal field generation process becomes more efficient. This makes the reversal of the field faster. This effect in the pumping-dominated regime overpowers the increase of the cycle period due to a decrease in meridional flow speed.

\begin{figure}
\centering
\includegraphics[scale=0.4,width=8.6cm]{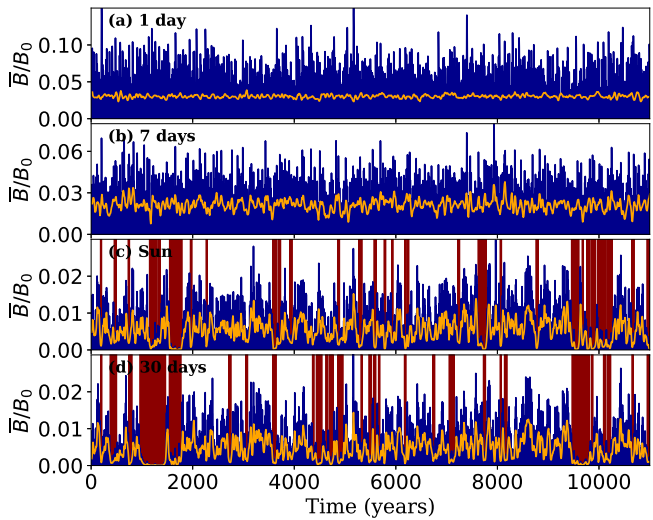}
\vspace{-0.2in}
\caption{Time series plot along with its smoothed variation of toroidal magnetic field for Model I of stars having rotation period of (a) 1 day, (b) 7 days, (c) 25.38 days (the solar value), and (d) 30 days. The dark-red bars highlight the extended weaker activity episodes i.e., grand minima in each case.}
  \label{fig:sin_series}
\end{figure}

\begin{figure}
\centering
\includegraphics[scale=0.43]{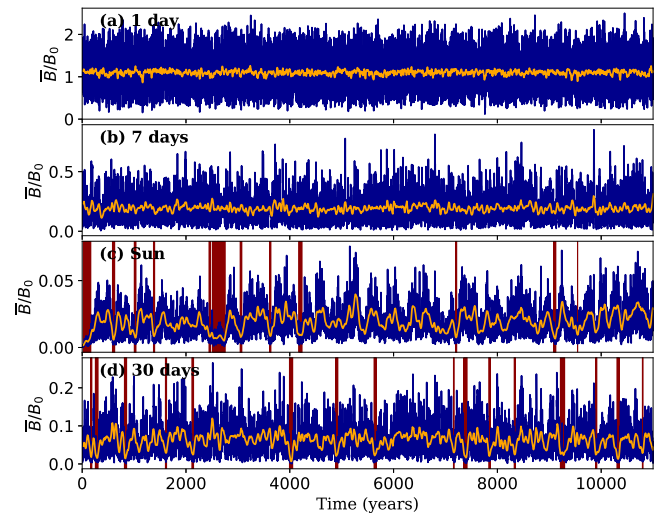}
\vspace{-0.2in}
\caption{Same as \Fig{fig:sin_series} but computed from 
the absolute radial magnetic field, averaged over the whole surface for Model III.
}
  \label{fig:hazra_series}
\end{figure}

\begin{figure}
\centering
\includegraphics[scale=0.38]{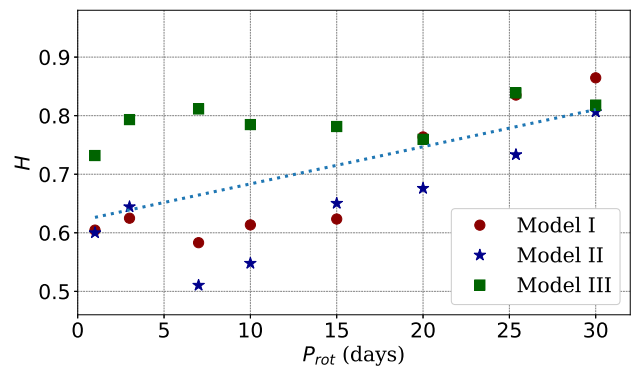}
\vspace{-0.3in}
\caption{Variation of Hurst exponent with respect to the rotation rate along with the linear-fit curve of all the models.}
\label{fig:hurst}
\end{figure}

\begin{figure*}
\centering
 \includegraphics[scale=0.38]{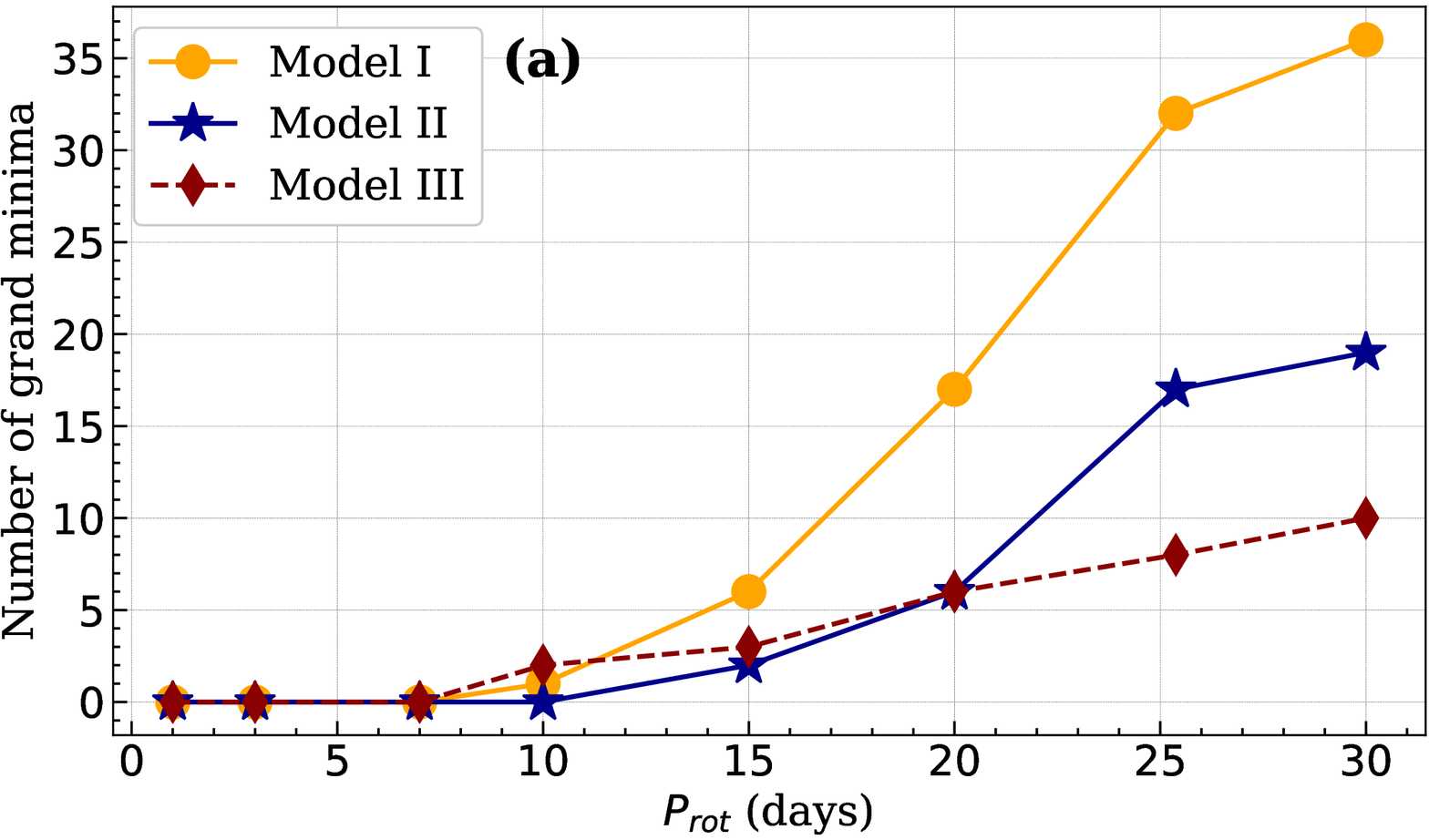}
 \includegraphics[scale=0.38]{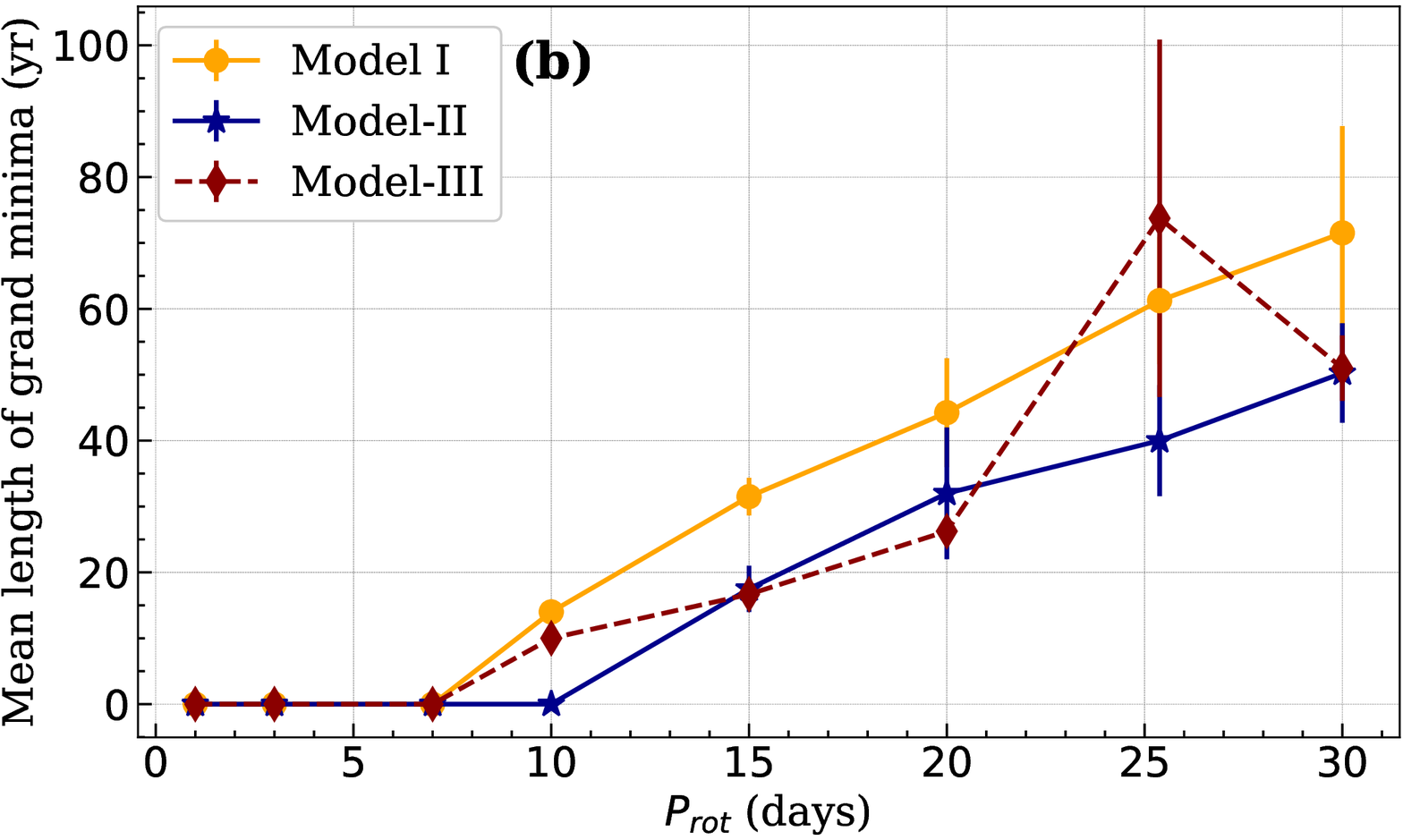}
 \vspace{-0.1in}
 \caption{Change of (a) the number and (b) the average duration of grand minima with the rotation period of stars. Yellow circles, blue asterisks, and red diamonds depict the trends for Models I, II, and III, respectively. In (b), the error bars are computed from the standard deviation of the durations of the grand minima in each case.
 }
  \label{fig:grand_min}
\end{figure*}

\subsection{Variability and grand minima occurrence}
\label{sec:variability}
We now come across the central question of our study %i.e., 
, i.e. how the long-term variability of the stellar cycles changes with the rotation rates of the stars. 
In our model, the cycle variability is produced due to the randomness in the Babcock–Leighton mechanism.
To analyze the long-term variability qualitatively, we carefully observe the time series data of the toroidal magnetic flux of the northern hemisphere and the absolute radial magnetic field averaged over the whole surface from a simulation of 11,000 years.
\Figs{fig:sin_series}{fig:hazra_series} show the discussed time series plots of the toroidal flux for Model~I and the radial field from Model~III.

From these figures, we can easily see that the fast rotators produce irregular cycles with smaller long-term variability, and on the other hand, slowly-rotating stars produce more long-term modulation in their cycles with episodes of extended weak magnetic fields.

To make a quantitative estimate of the irregularities and the long-term memory in the time series for each case, 
we compute the well-known Hurst exponent ($H$), which gives a measure of the temporal memory or persistence in the time series. A value of $H = 0.5$ implies that the time series is obtained from a memoryless random process. On the other hand, if a time series gives $H > 0.5$, then it suggests to have persistence. 
When a system has a memory that depends on the previous step, it is said to be persistent and thus in the time series 
if there was an increase in the value, it is more likely that the following step(s) will increase as well. In this case, the time series will cover more ‘distance’ than a random walk can. When $H<0.5$, the opposite applies and it is said to be anti-persistent.
When there is a long-term modulation in the stellar cycle data, we expect a memory and the value of $H$ should be larger than 0.5. A larger value of $H$ implies large long-term memory in the stellar cycle.

To obtain the Hurst exponent, we use the famous $R/S$ method as given in
\citet{ManWal69} and applied to many solar data in the past \citep{Ruz94, Suyal9, Ratul22}.
To do so, we first bin the data by using a bin-size of half a year.
Then, to evaluate $H$, the binned time series is divided into several shorter time series of length $\tau = 50$. The average re-scaled range ($R/S$) is then calculated for each temporal window $\tau$. At last, the slope of the $log(R/S)$ vs $log(\tau)$ values gives the value of $H$.

\Fig{fig:hurst} shows the values of $H$ evaluated in all the models for each star. A linear fit to all the data shows the overall increase in $H$ with the rotation period.
After analyzing this figure, we come to the conclusion that for the rapidly rotating stars, there is little long-term modulation and cycles are more irregular. Whereas a long-term memory is seen in slowly-rotating stars. Hence, the persistence increases as the rotation period increases. 
Interestingly, these results are in-tune with the observations \citep{Baliu95,Olah16}.

Finally, we identify the grand minima from the time series of the toroidal field at the base of CZ and the surface radial field. For this, we employ the same method as used in \citet{USK07} for the Sun; i.e., we first bin the data by using a  bin-size of the duration of one cycle, then we filter the data by using Gleissberg’s low-pass filter 1-2-2-2-1. This gives us the smoothed data. Finally, the portion of this data that falls below 50$\%$ of its mean for at least two consecutive cycle periods, is considered as a grand minimum. Later we count these numbers of grand minima to evaluate the frequency of occurrence of grand minima in each case. The computed number of grand minima for all the models are listed in Table \ref{tab:table1}.

From \Fig{fig:grand_min}a, we infer that in all the models, the number of grand minima increases with the increase in rotation period. Rapidly rotating stars hardly produce any grand minima, in fact, stars with a rotation period of 7 days or less, do not produce any Maunder-like grand minima. On the other hand, slowly-rotating stars produce some grand minima with an increasing trend with the rotation period. This is because, with the increase of rotation period, the supercriticality of the dynamo decreases, and the dynamo is more prone to produce extended grand minima in this regime.
This result is as per \citet{Vindya21} where we observe a decrease in the frequency of occurrence of grand minima as the supercriticality increases.
To check the robustness of this result, we run our simulations with another set of Gaussian random numbers having the same mean and $\sigma$ as the previous case. We again find the same conclusion that stars with  $P_{cyc}\leq 7$ days, do not produce grand minima, and the number of grand minima increases with the increase in rotation period.

\begin{figure}
  \centering
   \includegraphics[scale=0.30]{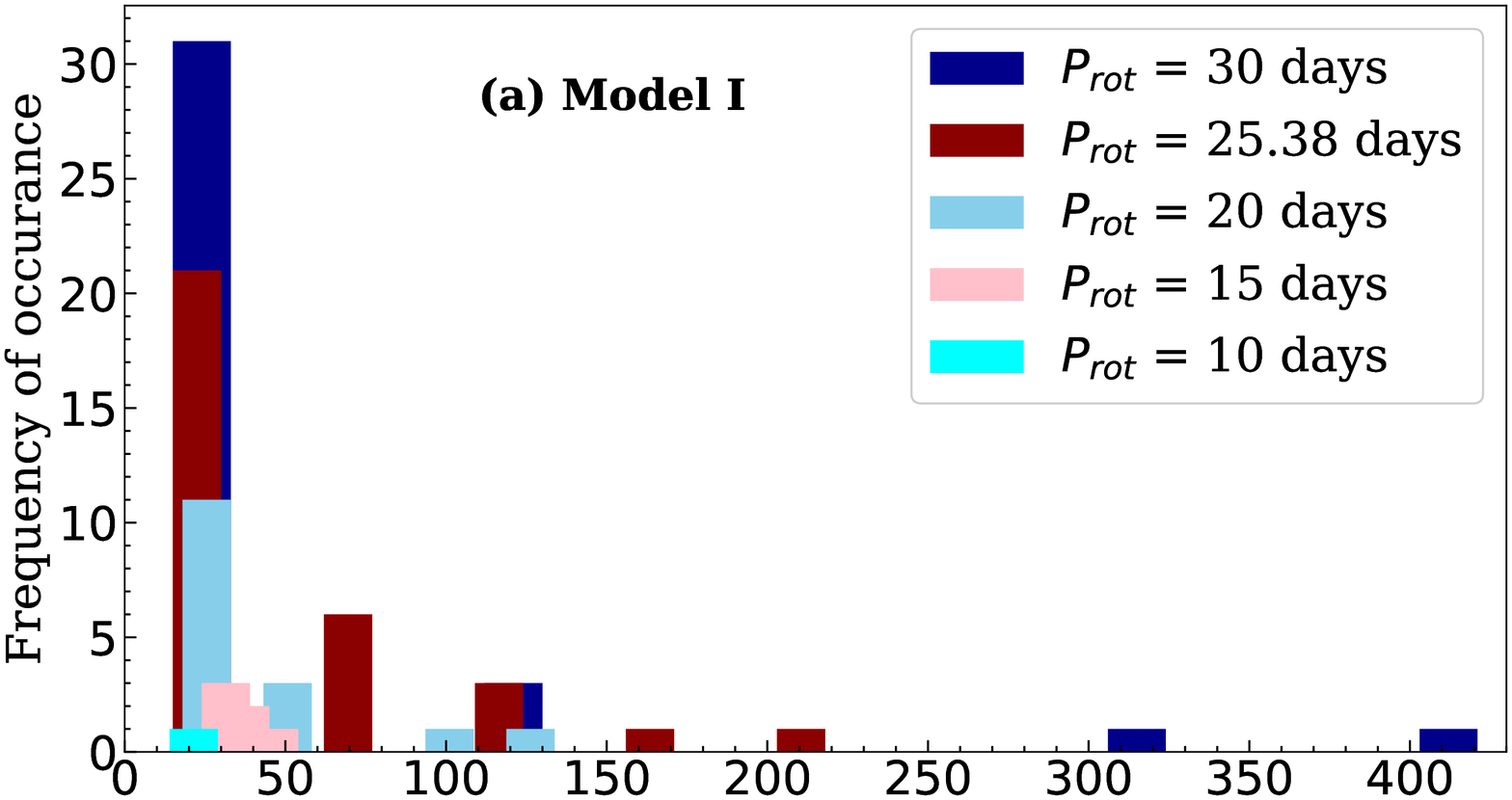}
   \includegraphics[scale=0.310]{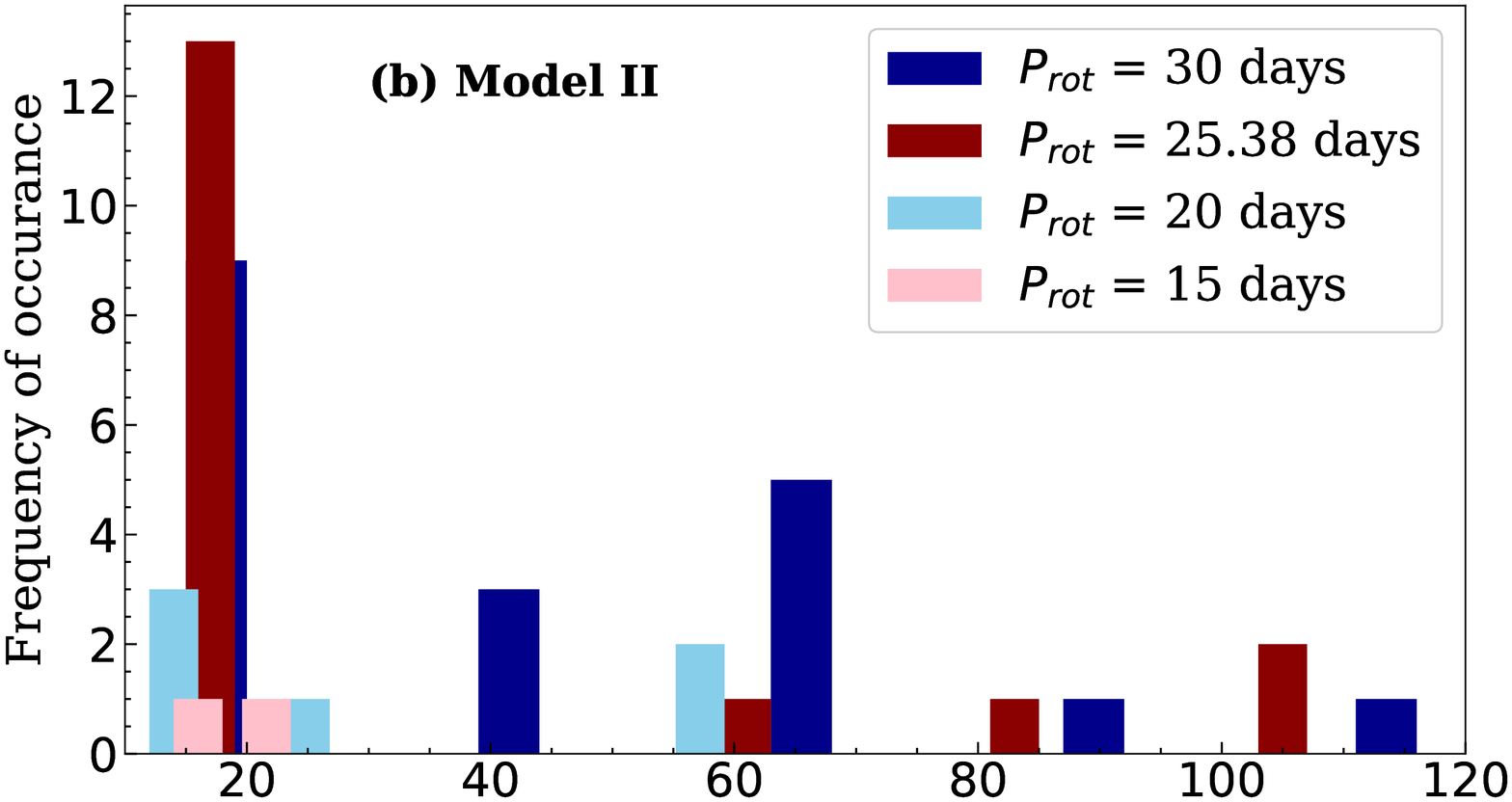}
   \includegraphics[scale=0.30]{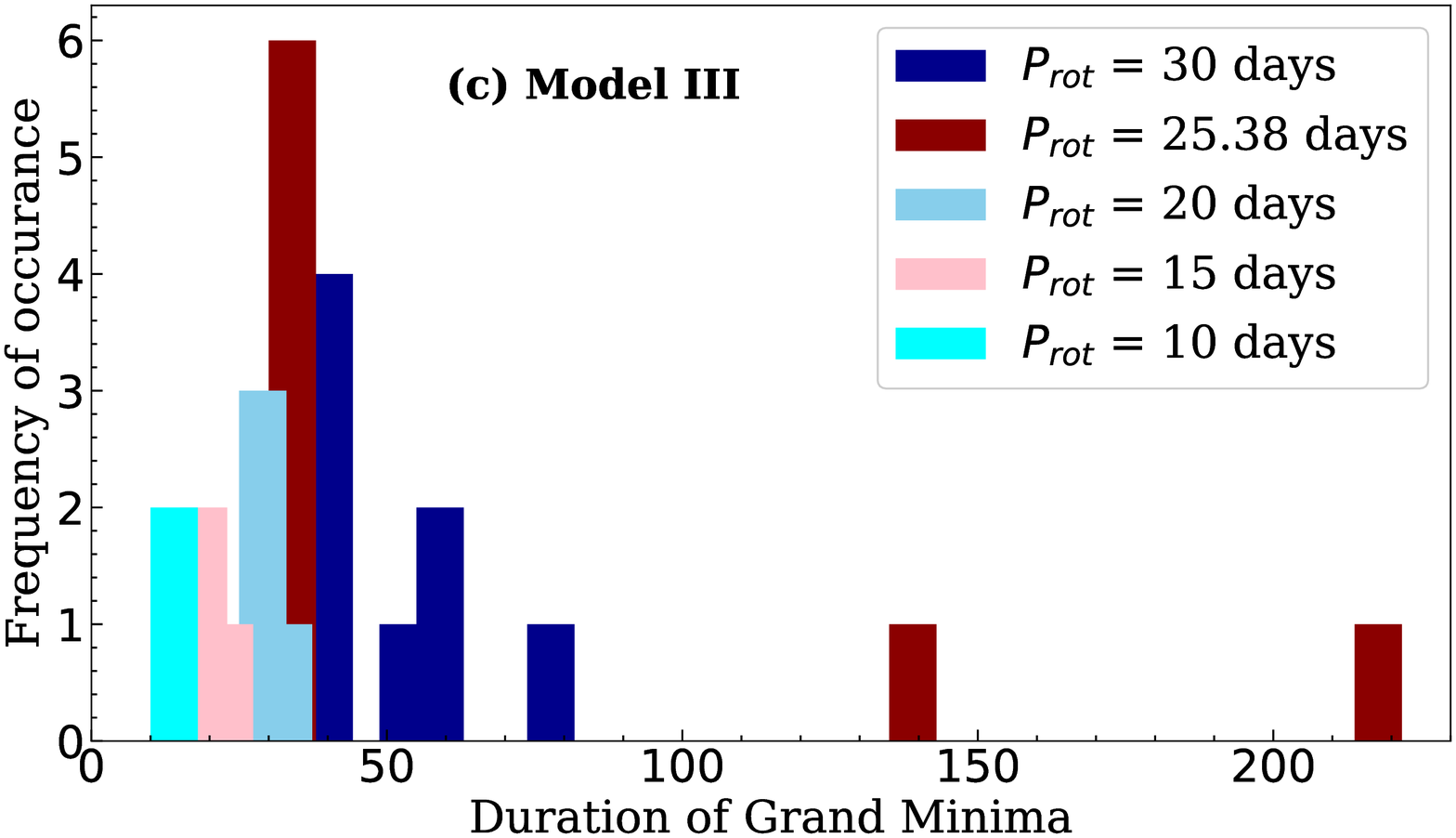}
%   \vspace{-0.2in}
  \caption{Relation between frequency of occurrence of grand minima with the corresponding duration in Model I-III for the stars in which grand minima are observed.} 
  \label{fig:hist}
\end{figure}

Additionally, we estimated the 
change in the average duration of grand minima with the stellar rotation. 
This variation is depicted in
\Fig{fig:grand_min}b.
Similar to the frequency, the average duration of grand minima increases with the increase of rotation period for all models except for the rotation period of 30 days in Model~III.
The association of the frequency of occurrence of the grand minima with the duration of these events is shown via histogram in \Fig{fig:hist}. With the help of \Fig{fig:grand_min}b and \Fig{fig:hist}, we can easily infer that the duration of the stellar grand minima falls mostly below 150 years, while the average value lies below 70 years only. 
In Model~I, we see a few grand minima occurring for longer duration due to the absence of hemispheric coupling. In full sphere models (Models II-III), hemispheric coupling helps to recover the model from extended grand minima easily \citep{KM18} and thus this does not produce very long grand minima. 
Further, in the Sun, we get about 10--40 grand minima (depending on which model we are considering), while in observations, this number is 27. 
And the result, that most of the solar grand minima hover below 150 years is in agreement with observations \citep{Uso17}.

\section{Conclusions and Discussion}
From our extensive simulations of the kinematic flux transport dynamo model with stochastically forced \bl\ source for the stars of 1M$_\odot$ mass with rotation periods of 1, 3, 7, 10, 15, 20, 25.38 (solar value), and 30 days, we make the following inferences.\\
(i) Rapidly rotating stars produce a strong magnetic field and the strength of the field increases with the increase of rotation rate which is in accordance with the observations \citep{Noyes84a}.
The increase of field with the increase of rotation rate in our model is due to the enhancement of the strength of the \bl\ source.\\
(ii) The cycle period increases with the increase of the rotation rate of the stars (in Models I-II) due to the weakening of the meridional circulation. However, when the downward turbulent magnetic pumping is included (in Model~III), cycles become longer in slowly-rotating stars and shorter (although become very irregular) in rapid rotators. Thus, pumping helps to bring the results closer to observations \citep{BoroSaikia18} as also suggested by \citet{Hazra19}.\\
(iii) Strong hemispheric asymmetry is produced in the magnetic field for all the stars. In general, the quadrupolar field dominates in the rapidly rotating stars and the dipolar field dominates in the sun and slowly-rotating stars.\\
(iv) In rapidly rotating stars, the stellar magnetic cycles are highly irregular, while in slowly-rotating stars cycles are more regular and the cycle amplitude displays a smooth long-term modulation. 
These results are consistent with the stellar observations \citep{Baliu95, BoroSaikia18, Olah16, garg19}.\\
(v) Only slowly-rotating stars with rotation period $\geq 10$~days produce grand minima. 
The number and the average duration of grand minima increase with the increase of the rotation period of the stars. This is again supported by the available observations because the confirmed Maunder minimum candidates are only slow rotators \citep[Sun ($P_{\rm rot}   =25.38$~d), HD 166620 ($P_{\rm rot} =45$~d);][]{anna2022}.\\
(vi) The length of the stellar grand minima lies mostly below 150 years. However, in the one hemisphere (Model~I) model, several grand minima occur with longer duration. The average duration of grand minima in this model is longer than the other two models because the hemispheric coupling is absent in this model. In full sphere models (Models II-III), hemispheric coupling helps to recover the model from extended grand minima easily \citep{KM18, HN19} 
and thus this does not produce very long grand minima. The result that most of the solar grand minima hover below 150 years is in agreement with the reconstructed solar activity data \citep{Uso17}.

Although many results of stellar cycles are robust and congruous with observations, there are limitations to our study. 
First, we have considered the only nonlinearity through the standard $\alpha$ quenching and
we have ignored the nonlinear feedback of the magnetic field on large-scale flows. While, in the sun, this is not a concern, in the rapidly rotating stars having a strong magnetic field, this nonlinearity can have a serious impact in producing cycle irregularity. 
Second, the turbulent transport coefficients are expected to change 
with the rotation and the magnetic field \citep{KPR94, Kar14b} and they can change the dynamo properties. 
Due to limited knowledge of their variations in different stars, we have not changed their values in our models. Third, the level of stochastic noise is kept constant in all the stars, again due to its limited knowledge.
Fourth, we have not considered the turbulent $\alpha$ effect, which in the Sun is negligible in comparison to the \bl\ process \citep{CS15}, but may becomes increasingly important in the rapidly rotating stars. 
However, our results of the trend of the 
cycle variability and the grand minima are not expected to change with many details of the model (e.g., type of nonlinearity, stochastic fluctuations, turbulent transport) because they depend on the amount of supercriticality of the model. 
It is obvious to accept that the dynamo supercriticality decreases with the decrease of the rotation rate of the star (mainly due to the decrease of $\alpha$).
Extended grand minima are easy to produce when the dynamo is near critical \citep{KO10, Vindya21}.
Also, observations hint that the solar dynamo (which produce grand minima) is operating near the critical transition \citep{R84, Met16}.
 Furthermore, observing some robust results (mainly the increase of the number and duration of grand minima with the increase of rotation period) in all the models having different parameters, we can have some confidence that our results will be validated in the more realistic stellar dynamo models and observations.

\section*{Acknowledgements}
We express our gratitude to Ricky Egeland for encouraging us to work on this problem. Additionally, we would like to acknowledge the contributions of Allan Sacha Brun and the anonymous referee, who provided valuable feedback and raised insightful questions, which enhances the quality of the paper.
Financial Support from the Department of Science and Technology (SERB/DST), India, through the Ramanujan fellowship (project No. SB/S2/RJN-017/2018) awarded to B.B.K. is acknowledged.
V.V. acknowledges the financial support from the DST through INSPIRE fellowship. 
L.K. acknowledges financial support from the Ministry of Science and High Education of the Russian Federation. 

\section*{Data Availability}
In our work, the dynamo calculations are done using a freely available code: $Surya$  \citep{NC02,CNC04}. Data from our dynamo models and the analysis codes can be shared upon a reasonable request.
%%%%%%%%%%%%%%%%%%%%%%%%%%%%%%%%%%%%%%%%%%%%%%%%%%



\bibliographystyle{mnras}
\bibliography{paper} % if your bibtex file is called example.bib

%\begin{thebibliography}{99}
%\end{thebibliography}

%%%%%%%%%%%%%%%%%%%%%%%%%%%%%%%%%%%%%%%%%%%%%%%%%%

%%%%%%%%%%%%%%%%% APPENDICES %%%%%%%%%%%%%%%%%%%%%

%\appendix


%%%%%%%%%%%%%%%%%%%%%%%%%%%%%%%%%%%%%%%%%%%%%%%%%%

% Don't change these lines
\bsp	% typesetting comment
\label{lastpage}
\end{document}